\documentclass[10pt,conference]{IEEEtran}

\usepackage{booktabs,balance} 
\usepackage{caption,subcaption,algorithm,algorithmic,tabularx,enumitem}
\usepackage{diagbox,multirow,graphicx,array,color}

\usepackage[hyphens]{url}
\usepackage{hyperref}
\usepackage[hyphenbreaks]{breakurl}
\usepackage[english]{babel}
\usepackage{blindtext}

\newcolumntype{Y}{>{\centering\arraybackslash}X}

\newcolumntype{L}[1]{>{\raggedright\let\newline\\\arraybackslash\hspace{0pt}}m{#1}}
\newcolumntype{C}[1]{>{\centering\let\newline\\\arraybackslash\hspace{0pt}}m{#1}}
\newcolumntype{R}[1]{>{\raggedleft\let\newline\\\arraybackslash\hspace{0pt}}m{#1}}

  \author{\IEEEauthorblockN{M. Hammad Mazhar}
    \IEEEauthorblockA{\textit{Department of Computer Science.} \\
      \textit{The University of Iowa}\\
      Iowa City, United States of America \\
      muhammadhammad-mazhar@uiowa.edu}
    \and
        \IEEEauthorblockN{Zubair Shafiq}
    \IEEEauthorblockA{\textit{Department of Computer Science.} \\
      \textit{The University of Iowa}\\
      Iowa City, United States of America \\
      zubair-shafiq@uiowa.edu}

    }

\begin{document}
\title{Characterizing Smart Home IoT Traffic in the Wild}
\maketitle

\begin{abstract}
As the smart home IoT ecosystem flourishes, it is imperative to gain a better understanding of the unique challenges it poses in terms of management, security, and privacy.
Prior studies are limited because they examine smart home IoT devices in testbed environments or at a small scale.
To address this gap, we present a measurement study of smart home IoT devices \textit{in the wild} by instrumenting home gateways and passively collecting real-world network traffic logs from more than 200 homes across a large metropolitan area in the United States.
We characterize smart home IoT traffic in terms of its volume, temporal patterns, and external endpoints along with focusing on certain security and privacy concerns.
We first show that traffic characteristics reflect the functionality of smart home IoT devices such as smart TVs generating high volume traffic to content streaming services following diurnal patterns associated with human activity.
While the smart home IoT ecosystem seems fragmented, our analysis reveals that it is mostly centralized due to its reliance on a few popular cloud and DNS services. 
Our findings also highlight several interesting security and privacy concerns in smart home IoT ecosystem such as the need to improve policy-based access control for IoT traffic, lack of use of application layer encryption, and prevalence of third-party advertising and tracking services.
Our findings have important implications for future research on improving management, security, and privacy of the smart home IoT ecosystem.
\end{abstract}


%

%
%

\section{Introduction}
Smart home IoT devices are used for a variety of home  monitoring and automation tasks such as smart locks and door bells, temperature and moisture sensors, and smart speakers for home assistance or streaming music.
The smart home IoT market has seen rapid growth over the past few years.
More than 832 million smart home IoT devices are expected to ship worldwide in 2019 \cite{smarthomeship2019IDC}.
Smart home IoT devices connect to the Internet to perform many of their tasks, such as accessing weather reporting services for home environment control and accessing media streaming services for providing entertainment.
Perhaps unsurprisingly, IoT traffic is now a major contributor to the overall Internet traffic. 
IoT traffic is expected to account for more than half of the Internet traffic by 2022.
48\% all IoT traffic is expected to be contributed by smart home IoT devices by 2022 \cite{ciscoVNI}.

The proliferation of smart home IoT has brought about many challenges such as management (e.g. device identification \cite{sivanathan2018tmc,ortiz2019iotdi}),  security (e.g. Mirai botnet \cite{krebsMirai, antonakakis17usenixsec}), and privacy (e.g. IoT devices leaking sensitive information  \cite{wood2017iotsnp,chu2019iot}).
Tackling these challenges drives research into understanding how smart home IoT devices are designed, adopted, and used.
However, conducting this research brings its own set of challenges.
\textit{First}, the smart home IoT ecosystem is fragmented with a wide variety of devices that are generally not amenable to inspection through standardized interfaces.
To overcome this challenge, we leverage the home gateway as the universal vantage point to inspect the network traffic generated by smart home IoT devices without needing to individually instrument them.
\textit{Second}, the behavior of smart home IoT devices is dependent on the environment they are placed in.
While smart home IoT devices may be studied in controlled testbed environments \cite{ren2019imc,wood2017iotsnp,alrawi19snp,sivanathan2017smartcity,moghaddam2019watching}, it may not reflect their real-world behavior.
Therefore, we study smart home IoT devices \textit{in the wild} through our home gateway instrumentation.
This allows us to capture real-world smart home IoT device behavior.
\textit{Finally}, studying smart home IoT behavior at scale is burdensome.
The diversity in the smart home IoT market in terms of the types of devices and manufacturers makes it difficult for researchers to gain insights or propose solutions applicable to the broader smart home IoT ecosystem.
We capture this diversity and scale by recruiting more than 200 homes to install our instrumented gateways and collect network traffic logs of smart home IoT devices \textit{in situ}.
Our logs contain network traffic from 1,237 devices including 66 different types of smart home IoT devices spanning categories such as smart assistants, smart TVs, and smart cameras. 
To protect privacy of users, we anonymize any personally identifiable information (e.g. IP addresses) and do not collect packet payloads in our network traffic logs.

Our analysis of smart home IoT traffic in the wild highlights three main characteristics:

\begin{itemize}[leftmargin=*]

     \item \textit{Device functionality} drives how much, when, and with whom smart home IoT devices communicate; media functionality generates high volume traffic, device traffic time series exhibit diurnal human activity patterns, and Internet services related to device functionality (e.g. video streaming services for smart TVs and online gameplay services for game consoles) generate most traffic.
     By understanding these behaviors, operators can better manage IoT devices on their networks such as by suitably provisioning interconnects to cloud networks hosting IoT back-ends. 
     
    \item While the smart home IoT ecosystem seems fragmented on the front-end, it is increasingly centralized on the back-end.
    Back-ends for smart home IoT devices are typically hosted on a few major cloud providers such as Google Cloud and Amazon AWS.
    These two account for 60-90\% of traffic for smart TVs,  smart speakers, smart assistants, and home automation devices.
    Smart home IoT devices are often configured with hard-coded DNS servers such as Google public DNS.
    98\% of smart assistants and 72\% of smart TVs use hard-coded Google DNS servers to resolve DNS queries instead of using the default DNS server configured at the home gateway.
    
    \item Smart home IoT devices present serious privacy issues because of their lack of use of traffic encryption and susceptibility to user behavior tracking.
    Some smart home IoT devices still communicate over (plain) HTTP, which leaves their traffic trivially vulnerable to eavesdropping and manipulation by network adversaries.
    20\% of smart assistant, smart TV, and health and wearable traffic is sent over HTTP.
    We also observe that several smart home IoT devices communicate with well-known third-party advertising and tracking services, complementing prior work \cite{razaghpanah2018apps}.
    5.9\%, 3.1\%, and 2.9\% of the hostnames accessed by smart TVs, game consoles, and smart assistants respectively were associated with known advertising and tracking services.
\end{itemize}

\noindent \textbf{Paper Organization:} The rest of the paper is set as follows.
We provide a brief background of the smart home IoT ecosystem, discuss our instrumentation for data collection, and present our dataset in Section \ref{sec:background}.
Section \ref{sec:characterization} presents the our characterization of smart home IoT traffic in the wild followed by a study on security and privacy issues in smart home IoT in Section \ref{sec:case_studies}.
We then discuss related work in Section \ref{sec:related_work} before concluding in Section \ref{sec:conclusion}.

\section{Background \& Data Collection}\label{sec:background}
\subsection{Background}

\begin{figure}
    \centering
    \includegraphics[width=\columnwidth]{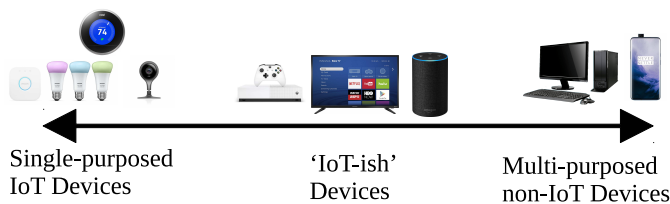}
    \caption{Spectrum of Internet-connected devices.}
    \label{fig:iot_spectrum}
\end{figure}

The proliferation of `smart' Internet-connected devices that can be remotely accessed and controlled has lead to the coining of the term `Internet of Things' or IoT.
Of particular note are smart home IoT devices, such as light bulbs, thermostats, and TVs that are commonly found in a home but were traditionally not connected to the Internet.
Smart home IoT device shipments are expected to reach 832 million in 2019, to grow to 1.6 billion shipped devices in 2023 \cite{smarthomeship2019IDC}.
These smart home IoT devices lie on a spectrum of Internet-connected devices based upon their functionality.
On one end, there are single-purposed devices such as smart light bulbs and thermostats that are typically considered IoT.
On the other end, there are multi-purposed devices such as smartphones and laptops that are typically not considered to be IoT.
In between, there are `IoT-ish' devices such as smart TVs and game consoles that are multi-purposed but are closer to IoT devices based on their main/core functionality. 
Figure \ref{fig:iot_spectrum} illustrates these devices on the spectrum of Internet-connected devices.
For the purpose of this work, we refer to single-purposed home IoT and home IoT-ish devices as smart home IoT devices.

\begin{figure}[!t]
	\centering
	\includegraphics[width=\columnwidth]{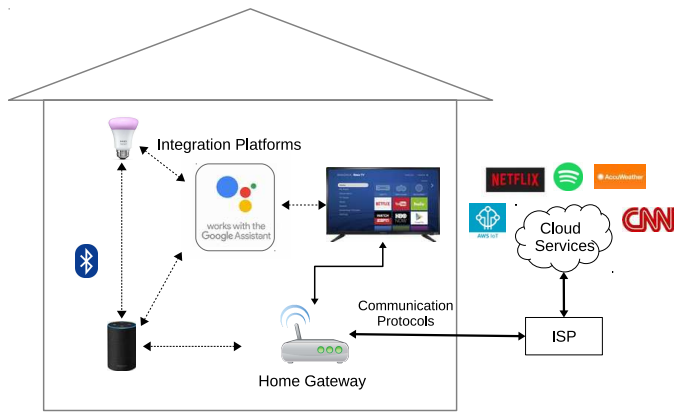}
	\caption{Overview of the smart home ecosystem. Smart home IoT devices can communicate with devices on the local network via different network technologies, coordinating actions via integration platforms. Connection to cloud-based services through communication protocols is mediated by the home gateway which provides Internet connectivity.}
	\label{fig:iot_home}
\end{figure}

Figure \ref{fig:iot_home} illustrates a typical smart home environment, outlining where each aspect of the smart home ecosystem lies.
The smart home ecosystem comprises of the various aspects (integration platforms, communication protocols, network technologies, cloud back-end services) developed to support IoT devices in a smart home.
Integration platforms such as Apple's HomeKit \cite{applehomekit2019}, Amazon's Alexa \cite{alexa2019}, Google's Home \cite{googlehome2019}, and Samsung's SmartThings \cite{smartthings2019} allow smart home IoT devices to implement their functionality in a coordinated manner (e.g. allowing a light sensor detecting low sunlight to turn on smart bulbs).
Smart home IoT devices use a variety of communication protocols such as HyperText Transport Protocol (HTTP/HTTPS), Message Queuing Telemetry Transport (MQTT), Domain Name System (DNS), and Universal Plug and Play (UPnP).
Smart home IoT devices also use physical-layer network technologies such as Zigbee \cite{zigbee2019}, Z-Wave \cite{zwave2019}, and Bluetooth Low Energy (BLE) \cite{bluetooth2019} for local communications,  and Wi-Fi or Ethernet for Internet connectivity.
Finally, smart home IoT devices also rely on cloud back-end services for data storage and backup, firmware updates, remote access and integration, and other services for media streaming, weather updates, and news reports.

\subsection{Data Collection}
The home gateway provides a central vantage point to measure the characteristics of all  devices in a smart home.
We can passively monitor network traffic generated by smart home IoT devices in smart homes as they connect to their cloud services and other third-party services on the Internet.
In this section, we discuss the instrumentation of the home gateway for this task along with the challenges associated and the dataset collected via this vantage point.

\vspace{.05in} \noindent\textbf{Home gateway instrumentation.}\label{sec:data_collection_setup}
We partner with a home gateway management software company to utilize the home gateway.
The company provides a Web-based platform for smart home users to manage their home gateway, providing features such as Internet access control, device security and bandwidth management.
Off-the-shelf commodity gateway routers are instrumented with a modified version of OpenWRT, a Linux-based operating system for networking devices.
This instrumentation is designed to passively collect information on network traffic and the devices connected to the gateway router to provide the desired services.%

\vspace{.05in} \noindent\textbf{Network traffic data.}
As commodity gateway routers are typically limited in terms of processing power and memory, the instrumentation for collecting such information has to be lightweight to prevent negative impacts on the router's primary purpose of packet forwarding.
To this end, the home gateways are instrumented to collect flow-level summary information for network traffic instead of detailed packet-level header and payload information.
A flow is defined as a time-contiguous data transfer between two unique endpoints, where one endpoint lies on the local network and the other is external to the local network (e.g. on the Internet).
The home gateway maintains a table of such flows along with their summary information and uploads this table after a fixed time interval (30 seconds) to a secure cloud server designated for data collection, after which the table is flushed from memory.
It is noteworthy that no Deep Packet Inspection (DPI) is performed when collecting this data, so application-layer information (e.g. URLs) is not available, even when in cleartext.
The summary information includes data such as:
\begin{itemize}
    \item \textbf{External IP addresses.}
    \item \textbf{Hostname of the external IP address.} This is determined by querying the external IP address in the gateway's DNS cache.
    \item \textbf{Direction of flow.} Either to or from the local IP address.
    \item \textbf{Bytes Transferred.}
\end{itemize}

\vspace{.05in}  \noindent\textbf{Network device fingerprinting.}
The home network management platform also incorporates network device fingerprinting into its services, providing users with information regarding what devices connect to their network.
The home gateway is instrumented to upload Simple Service Discovery Protocol (SSDP), Dynamic Host Configuration Protocol (DHCP) and UPnP traffic to the cloud.
This traffic is then matched to expert rules crafted through analysis of such traffic by domain experts to identify devices.
This approach is similar to the expert rule generation approach outlined by Kumar et al. \cite{kumar18usesec}.
The user can cross-check this identification and inform customer support if it is incorrect, in which case the rules are updated to reflect the correct traffic-to-device mapping.
These rules may fail to correctly identify devices in cases where reported values in SSDP and DHCP traffic correspond only to the networking components employed by the devices, such as wireless chipsets.
We take into account such devices when we count the number of devices in a smart home, but do not study their behavior in further analysis.

\noindent \textbf{Ethical Considerations.} The company collects data from its customers for not only providing current services, but also for research and development purposes.
Data for the latter is collected from a special subset of users who have consented to the use of their data for this purpose.
These users include early adopters as well as friends and family of employees of the company.
We analyze \textit{anonymized data} from these users about smart home IoT device behavior.
We only use the flow-level summary information outlined in Section \ref{sec:data_collection_setup}, where personally identifiable information (such as MAC addresses of devices or public-facing IP addresses of homes) is not collected.
Individual devices and home gateways are anonymized using randomly generated IDs, so we can identify which devices are connected to which gateway, but do not identify who these gateways and devices belong to in the real world.
For each device in our dataset, we collect its device type as identified by the fingerprinting approach outlined earlier.

\begin{table*}[!ht]
	\begin{tabularx}{\textwidth}{|p{3.1cm}|R{0.8cm}|R{0.8cm}|X|R{0.8cm}|R{1.4cm}|R{1.5cm}|}
		\hline
		\textbf{Device Category}&\textbf{Device Count}&\textbf{Home Count}& \textbf{Manufacturers } &\textbf{Unique Device Types}&\textbf{Mean download per day per Device (GB)} &\textbf{Mean upload per day per Device (GB)} \\\hline
		Smart TV                    & 78 & 55 & Samsung, TCL, Vizio, LG, Sharp, Sony, Apple, Google, Roku, Arcadyan, LiteOn      & 29 & 3.53 & 0.06 \\\hline
		Game Console                & 45 & 38 &Nintendo, Microsoft, Sony                                                         & 8 & 3.7 & 0.1 \\\hline
		Smart Speaker               & 29 & 9 & Sonos, Russound                                                                   & 10 & 0.06 & 0.002\\\hline
		Smart Assistant             & 28 & 21 & Google, Amazon                                                                   & 2 & 0.3 & 0.01\\\hline
		Smart Camera                & 16 & 5 & Belkin, Netgear, Nest                                                             & 3 & 0.06 &1.2\\\hline
		Work Appliance              & 14 & 14 & Canon, Epson, Brother, HP                                                        & 8 & 0.0002 & 0.0005\\\hline
		Health \& Wearable          & 14 & 12 & Apple, Fitbit, Peloton                                                           & 3 & 0.0004 & 0.00009\\\hline
        Home Automation             & 16 & 5 & Control4, Nest, Phillips, Solarcity, iRobot, LAMetric                             & 7 & 0.001 & 0.002\\\hline\hline
		Smartphone                  & 473 & 173 & Samsung, Nokia, Motorola, Apple, LG, ASUS, HTC, Huawei, OnePlus, ZTE,          & 31 & 0.4 & 0.05\\\hline
		Computers/Laptops           & 372 & 148 & Apple, Intel, Microsoft, ASUS, Gigabyte, Samsung, HP, Lenovo, PC, Raspberry Pi & 9 & 0.3 & 0.1\\\hline
        Tablets                     & 95  & 62 & Amazon, Apple                                                                   & 3 & 0.5 & 0.1\\\hline
		Unknown                     & 32 & 27 & Xerox, Shenzen RF, China Dragon, Clover Network, Espressif                       & 28 & 0.25 & 0.26\\\hline
		Networking                  & 18 & 5 & Netgear, QNAP, TP-Link, Western Digital, Plume Design                             & 6 & 0.7 & 1.2\\\hline
		\textbf{13}                 & \textbf{1237} & \textbf{220 }&                                                             & \textbf{142} & &\textbf{Total}\\\hline		
	\end{tabularx}
	\vspace{0.05in}
	\caption{Basic statistics of smart home network-connected devices in our dataset. Devices are categorized based on their primary functionality. We consider the first 8 categories as \textbf{Smart Home IoT} devices.}
	\vspace{-0.2in}
	\label{tab:dev_types}
\end{table*}

\subsection{Data Statistics}

\noindent\textbf{Dataset.}
Our analysis is performed on data collected during a 19-day period in February 2018.
The data is collected from 220 homes spread across a large metropolitan area in the United States, with traffic from 1237 unique network-connected devices observed during data collection.
We break down these devices in terms of their functional categories, numbers and the amount of traffic they generated in Table \ref{tab:dev_types}.
We consider smart home IoT devices to include game consoles, smart TVs (including video streaming devices), smart speakers, smart assistants, smart cameras, work appliances, health devices \& wearables, and home automation devices.
Also, we consider smartphones, computers/laptops, networking devices and tablets as non-IoT devices.
Overall we observe 142 unique device types in our dataset, 66 of which we classify as smart home IoT devices and 48 as non-IoT devices.
The fingerprinting approach outlined previously was unable to identify 28 device types, which we label as Miscellaneous.
In all, we observed 240 smart home IoT devices, 958 non-IoT devices and 32 Miscellaneous devices in our dataset.
\begin{figure}[!ht]
    \centering
    \includegraphics[width=\columnwidth]{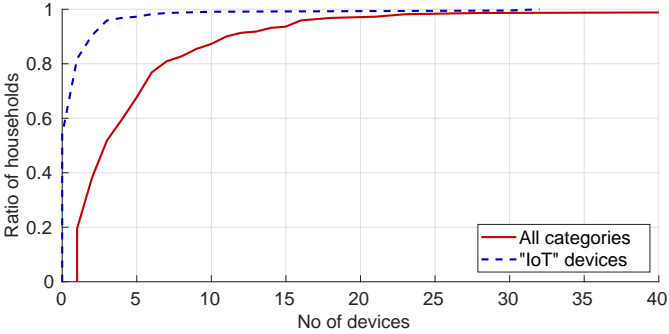}
    \caption{Distribution of device counts across homes in our dataset. We plot separate distributions when all device categories are considered and when only smart home IoT devices are considered.}
    \label{fig:house_dist}
\end{figure}

\noindent\textbf{Device distribution across homes.}
We first look at the distribution of the number of devices per home in our dataset in Figure \ref{fig:house_dist}, determined by the number of unique device IDs associated with each home, with separate distributions when all devices categories are considered and when only smart home IoT devices are considered.
We observe that around 51\% of homes had less than 3 devices connected directly to the instrumented gateway and 54\% of homes did not have a smart home IoT device connected directly to the instrumented gateway.
It is likely that such homes may have devices behind another networking device such as a Wi-Fi router masking their presence from our instrumented gateway, or they simply do not have many devices.
We however also observe a few homes with more than 50 devices and more than 25 smart home IoT devices connected to the gateway.
Our dataset covers a wide variety of homes which vary in terms of their adoption of `smart' home and is illustrative of the need to study smart home IoT in the wild.

\begin{figure}[!ht]
    \includegraphics[width=\columnwidth]{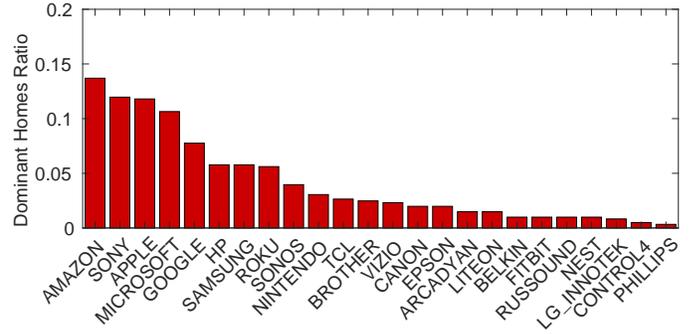}
    \caption{Manufacturer dominance in homes across smart home IoT devices.}
    \label{fig:manu_dom}
\end{figure}

\vspace{.05in} \noindent\textbf{Manufacturer dominance.}
Users may exhibit preferences for specific manufacturers when considering devices for their smart home, due to familiarity and ease of integration with other devices from the same manufacturer.
As such, we study whether there are any preferred or dominant manufacturers amongst the homes in our dataset.
We define a manufacturer to be dominant in a home if it has the highest amount of devices in the home or it is the only manufacturer in the home.
In cases where all devices belong to different manufacturers, we divide that home equally across all present manufacturers.
We present manufacturer dominance across smart home IoT devices in Figure \ref{fig:manu_dom}.
We observe 24 different manufacturers presenting some form of dominance in our dataset.
The most dominant manufacturer for smart home IoT devices is Amazon at 14\% of homes in our dataset, which produces the Echo line of home voice assistants and the Fire TV line of video streaming devices.
Next, Sony at 11\% of houses produces the Bravia line of smart TVs and the PlayStation line of consoles.
At par with Sony is Apple, which produces the Apple Watch wearable and the Apple TV amongst other smart home IoT devices, 
Moving further ahead, we see manufacturers of smart IoT device categories such as home automation devices (Google, Nest), smart speakers (Sonos, Russound), smart TVs and video streaming devices (Samsung, TCL, Vizio), and work appliances (Brother, Canon, Epson, HP). 
Given such diversity it becomes important to study smart home IoT devices in the wild, where insights can be considered more representative of how smart home IoT devices behave when used by real users.

\section{Smart Home IoT Activity in the Wild} \label{sec:characterization}
In this section, we discuss our analysis of smart home IoT device traffic.
We frame our analysis to ascertain whether the functionality provided by smart home IoT devices affects characteristics of their traffic.
To this end, our analysis answers three main questions:

%
\begin{itemize}
	\item \textit{How much do smart home IoT devices communicate over the Internet?} We shed light smart home IoT traffic volumes to illustrate how device functionality may affect device traffic volumes.
	
	\item \textit{When are smart home IoT devices communicating over the Internet?} By examining the temporal nature of smart home IoT traffic, we seek to understand if device functionality reflects in temporal traffic patterns.
	
	\item \textit{Who are smart home IoT devices communicating with?} By investigating whom different smart home IoT devices communicate with over the Internet, we seek to understand how device functionality determines what a device communicates with over the Internet.
	
\end{itemize}
\subsection{How much do smart home IoT devices communicate?}\label{sec:volume}

\vspace{.05in} \noindent \textbf{Traffic Volume.}
Table \ref{tab:dev_types} shows the average traffic volume per device per day for each device category.
We observe that smart home IoT devices such as smart cameras, game consoles, and smart TVs account for vastly more traffic volume than other categories because they download or upload media content. 
Game consoles, smart TVs download much more data than they upload, likely due to their main functionality to access media content.
Smart cameras upload much more data than they download as they are capable of uploading video footage.
Home automation, work appliances, and health and wearable devices account for less traffic volume because they only download or upload control traffic.

\begin{figure}
    \centering
    \includegraphics[width=\columnwidth]{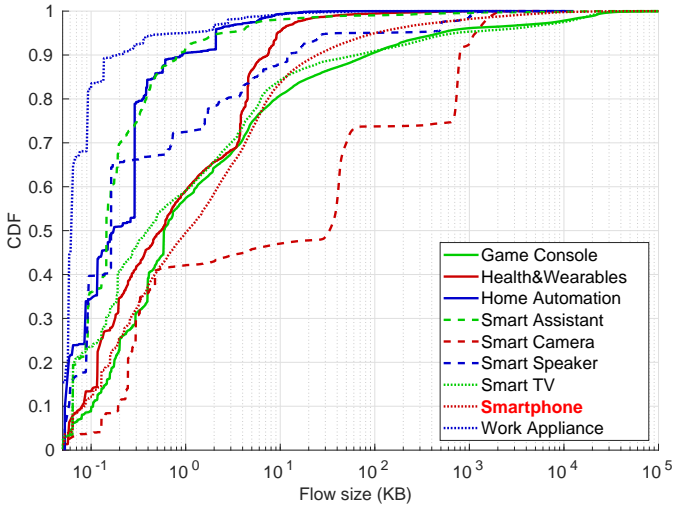}
    \caption{Flow size distributions for smart home IoT device categories, with smartphones as baseline. Some categories primarily less than a KB of traffic per flow, whiles others generate significantly more traffic per flow.}
    \label{fig:flow_dists}
\end{figure}

\vspace{.05in} \noindent \textbf{Flow Size.}
Figure \ref{fig:flow_dists} plots distributions of flow traffic sizes for smart home IoT device categories, with the distribution for smartphones as baseline.
We note that some device categories such as home automation, smart assistants and work appliances generate small flows. 
More than 85\% of the flows generated by these devices are less than one kilobyte. 
In comparison, smartphones have only 50\% of such small flows.
Smart TVs, game consoles, and health \& wearables exhibited similar flow distributions as smartphones.
Smart cameras generate large flows with over 25\% of flows more than a megabyte.
Such flows likely correspond to uploading of video footage for remote viewing and backup.

\vspace{.05in} \noindent \textbf{\textit{Takeaway.}}
\textit{Functionalities provided by smart home IoT devices play a pivotal role in the volume and flow size of the traffic they generate.
Devices that provide functionalities requiring high data volumes such as accessing Web content or uploading video data will generate high volumes of network traffic reflected in high-volume flows.
}
%

\begin{figure}[!ht]
    \centering
    \includegraphics[width=\columnwidth]{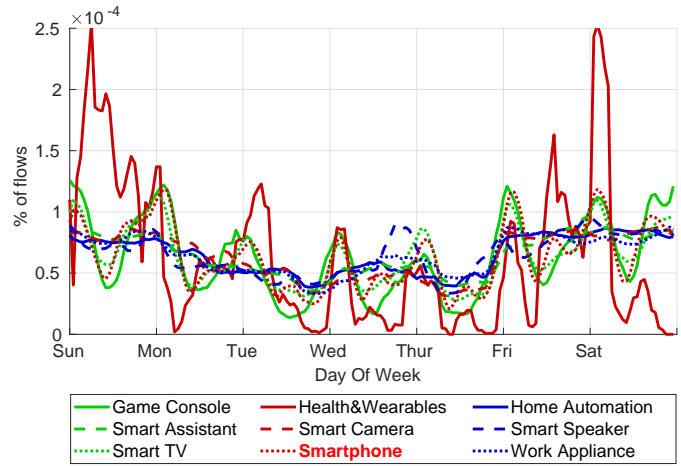}
    \caption{Smart home IoT device activity by category over the week, with diurnal and non-diurnal patterns. Devices with user-driven functionality exhibit diurnal activity patterns corresponding to human activity patterns.}
    \label{fig:daily_dist}
\end{figure}
\subsection{When do smart home IoT devices communicate?} \label{sec:time_comms}
We now look at smart home IoT device activity patterns to understand the temporal nature of their activity.

\vspace{.05in} \noindent \textbf{Diurnality.}
Figure \ref{fig:daily_dist} plots per-hour activity time series for IoT device categories over the course of a week, with smartphones as baseline.
We observe some device categories such as smart TVs, health and wearables, and game consoles exhibit a daily diurnal pattern that is driven by human activity patterns.
This diurnal pattern is characterized by lower activity in the middle of the day when people are expected to be at work, rising to a peak the end of the day when they return home.
We note device categories exhibiting such patterns have functionalities involving direct user interactions i.e. turning on the TV to watch video or using a game console to play games.
As a baseline, smartphones exhibit similar diurnal patterns.
We also observe that some other device categories do not exhibit such daily diurnal patterns, which illustrates that they are not dependent on user interactions.
For example, smart cameras and smart thermostats are designed to monitor home environments and as such remain equally active regardless of time of day.
For both groups, we observe higher device activity on the weekend (Friday-through-Sunday) than during the rest of the week because users are more likely to be at home during the weekend.
We specifically see a spike in traffic on the weekend for health \& wearable devices, which is due to the Peloton exercise bikes.
This fact further indicates that user presence in the home has an impact on device activity.

\begin{figure}
    \centering
    \begin{subfigure}[b]{\columnwidth}
    \centering
    \includegraphics[width=\columnwidth]{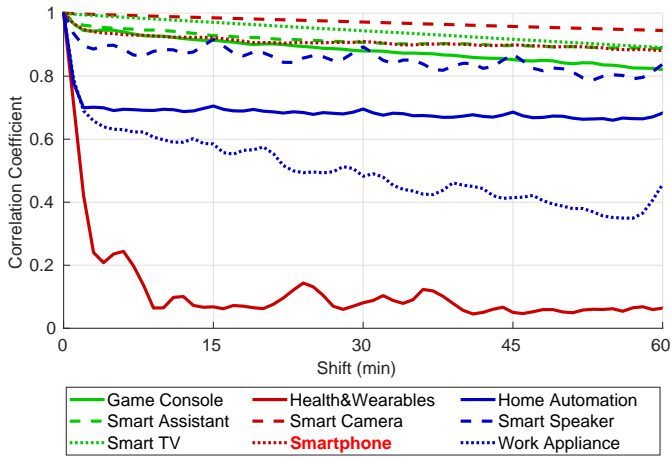}
    \caption{Outbound traffic}
    \label{fig:cross_out_periodic}
    \end{subfigure}
    \begin{subfigure}[b]{\columnwidth}
    \centering
    \includegraphics[width=\columnwidth]{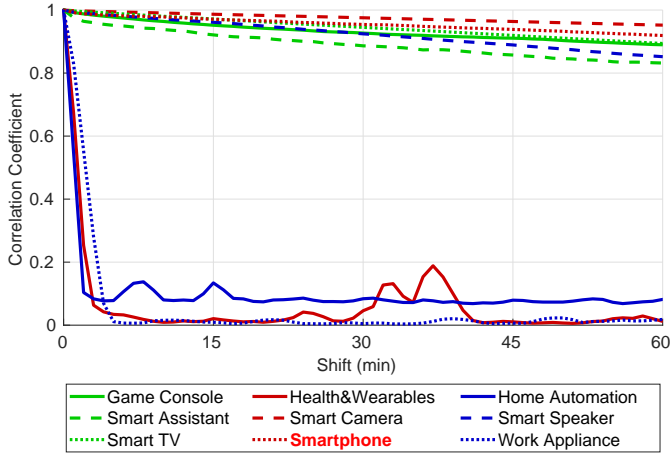}
    \caption{Inbound traffic}
    \label{fig:cross_in_periodic}
    \end{subfigure}
    \caption{Auto correlation coefficient distributions for smart home IoT device categories exhibiting periodicity. Devices generating programmed `heartbeat' traffic exhibit sub-hour periodicity.}
	\label{fig:cross_periodic_dist}
\end{figure}

\vspace{.05in} \noindent \textbf{Periodicity.}	
We compute the normalized auto correlation coefficient \cite{park2018autocorr} for per-minute activity time series.
Since some device categories did not exhibit any diurnality in their per-hour activity time series, we hypothesize that their per-minute activity time series may exhibit sub-hour periodicity.
The distribution of auto correlation coefficient over successive shifts can indicate the presence of periodic signals in the time series.
The length of the period is indicated by the distance between successive peaks in the auto correlation distribution.
Figure \ref{fig:cross_out_periodic} plots the auto-correlation distribution for outbound traffic.
We observe periods of 15 minutes for smart speakers and home automation services, and 1 hour for work appliances in outbound traffic.
However, this periodicity disappears when we consider inbound traffic in Figure \ref{fig:cross_in_periodic}.
Smart home IoT devices are often designed to generate outbound periodic `heartbeat' traffic that does not depend on user interaction.
As a baseline comparison, smartphones did not exhibit any sub-hour periodicity in traffic in either direction.

\vspace{.05in} \noindent \textbf{\textit{Takeaway.}}
\textit{Functionalities provided by smart home IoT devices determine their temporal activity patterns.
Devices with functionality requiring direct user interactions will exhibit daily diurnal patterns correlated with human activity patterns.
Devices with functionality not requiring direct user interactions may exhibit sub-hour periodicity due to ``heartbeat'' traffic.}

\subsection{Who are smart home IoT devices communicating with?}\label{sec:who}
To answer this question, we analyze smart home IoT device activity in terms of the network hosts they communicate with.
\begin{table}
    \begin{tabular}{p{1.3cm}| p{1.9cm} p{1.9cm} p{1.9cm}}
     \textbf{Category}& \textbf{AS 1 Org.}& \textbf{AS 2 Org.} &\textbf{AS 3 Org.} \\
     &\textbf{[\% of flows]}&\textbf{[\% of flows]}&\textbf{[\% of flows]}\\\hline
     Game & MICROSOFT & AMAZON-02 & AMAZON-AES \\
     Consoles & [26.6\%]         & [22.6\%]           & [10.7\%]\\\hline
    Smart   & GOOGLE & AMAZON-AES & AMAZON-02 \\
    TVs    & [46.8\%]         & [14.8\%]           & [11.0\%]\\\hline
    Smart & AMAZON-AES & AMAZON-02 & PANDORA \\
    Speakers              & [64.1\%]         & [16.8\%]           & [10.3\%]\\\hline
    Smart & AMAZON-02 & GOOGLE & AMAZON-AES \\
    Assistants & [64.1\%]         & [16.8\%]           & [10.3\%]\\\hline
    Smart & GOOGLE & AMAZON-AES & AMAZON-02 \\
    Cameras & [48.5\%]         & [45.7\%]           & [5.5\%]\\\hline
    Work  & HP-INTERNET & GOOGLE & TANDEM \\
    Appliance              & [91.5\%]         & [8.2\%]           & [0.1\%]\\\hline
    Health \& & APPLE-ENGINEERING & ERICYHOST & COMCAST \\
    Wearables & [63.5\%]         & [20.6\%]           & [4.5\%]\\\hline
    Home & GOOGLE & AMAZON-02 & AMAZON-AES \\
    Automation & [37.0\%]         & [25.2\%]           & [24.1\%]\\\hline
    
    \end{tabular}
    \caption{Top 3 ASes for each smart home IoT device category.}
    \label{tab:top_asns}
\end{table}

\noindent\textbf{Autonomous Systems.}
We first look at the Autonomous Systems (ASes) that smart home IoT devices communicate with.
We list the top 3 ASes by traffic for each smart home IoT device category in Table \ref{tab:top_asns}.
%
%
%
%
We observe specific organizations for specific categories, such as Microsoft for game consoles, HP for work appliances, Pandora for smart speakers, and Apple and Comcast for health and wearable devices.
Such organizations provide specific services such as online gameplay via Xbox Live for Microsoft or music services by Pandora. 
However, we also note that nearly all categories have their top ASes belong to either Google or Amazon, often accounting for 70-90\% of all traffic for the category. 
%
%
%
Both organizations provide general-purpose cloud services such as Amazon Web Services (AWS) and Google Cloud.
Devices manufactured by either company such as the Amazon Echo or the Google Chromecast would be expected to leverage these cloud services.
However other manufacturers also opt for these services to avoid setting up their own due to cost and efficiency issues.
For instance, Belkin uses AWS to provide cloud services for their Wemo line of products \cite{case2019belkin}.
While the smart home IoT ecosystem may seem heterogeneous from the diversity of manufacturers and products available, there is a \textit{centralization} of service delivery for smart home IoT, where most services are being provided through Google Cloud or AWS.
\begin{figure*}[!ht]
    \centering
    \begin{subfigure}[t]{0.3\textwidth}
    \vskip 0pt
    \centering
    \includegraphics[width=\textwidth]{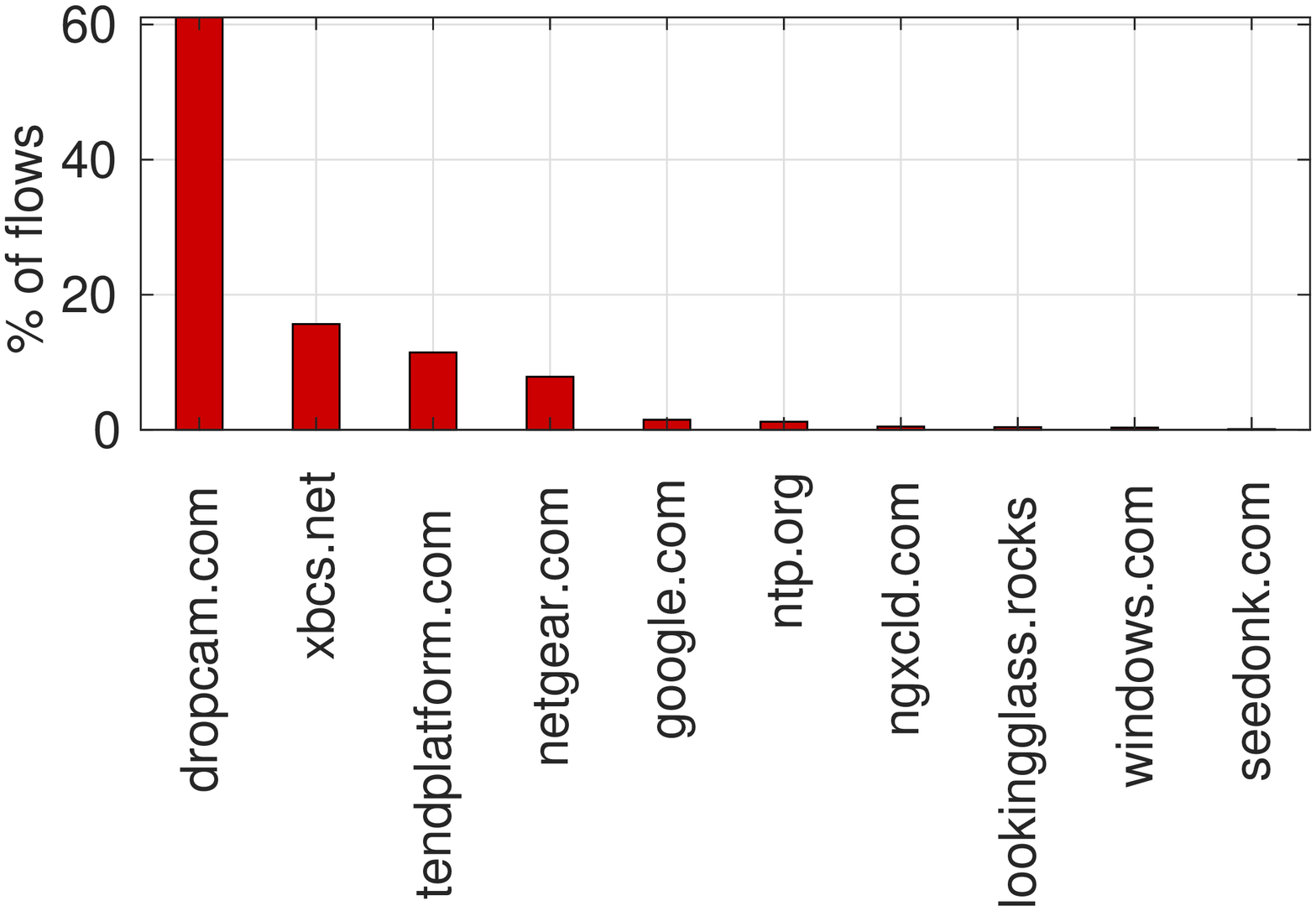}
    \captionsetup{skip=26pt}
    \caption{Smart Cameras}
    \label{fig:smart_camera}
    \end{subfigure}
    \hfill
    \begin{subfigure}[t]{0.3\textwidth}
    \vskip 0pt
    \centering
    \includegraphics[width=\textwidth]{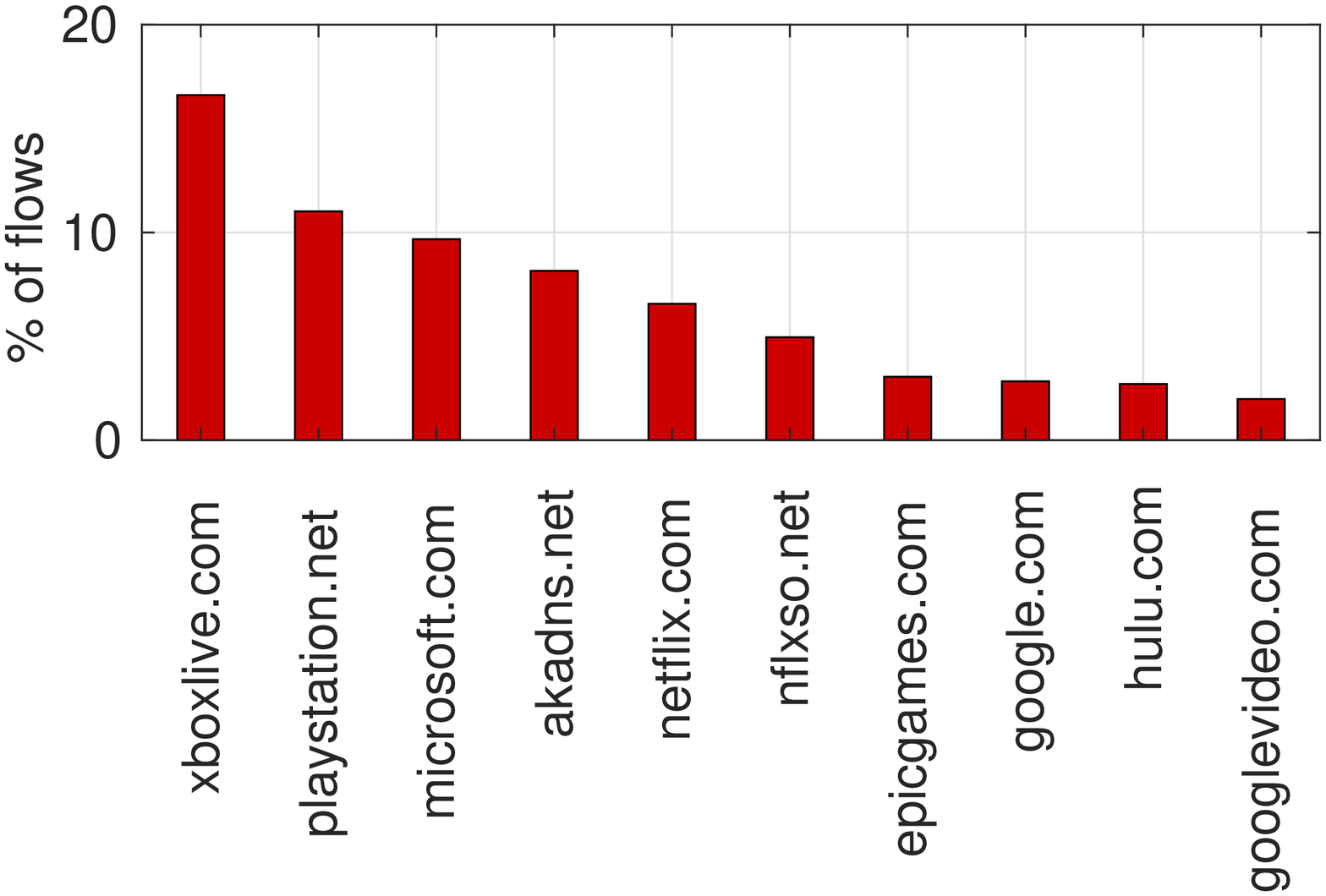}
    \captionsetup{skip=28pt}
    \caption{Game Consoles}
    \label{fig:game_domain}
    \end{subfigure}
    \hfill
    \begin{subfigure}[t]{0.3\textwidth}
    \vskip 0pt
    \centering
    \includegraphics[width=\textwidth]{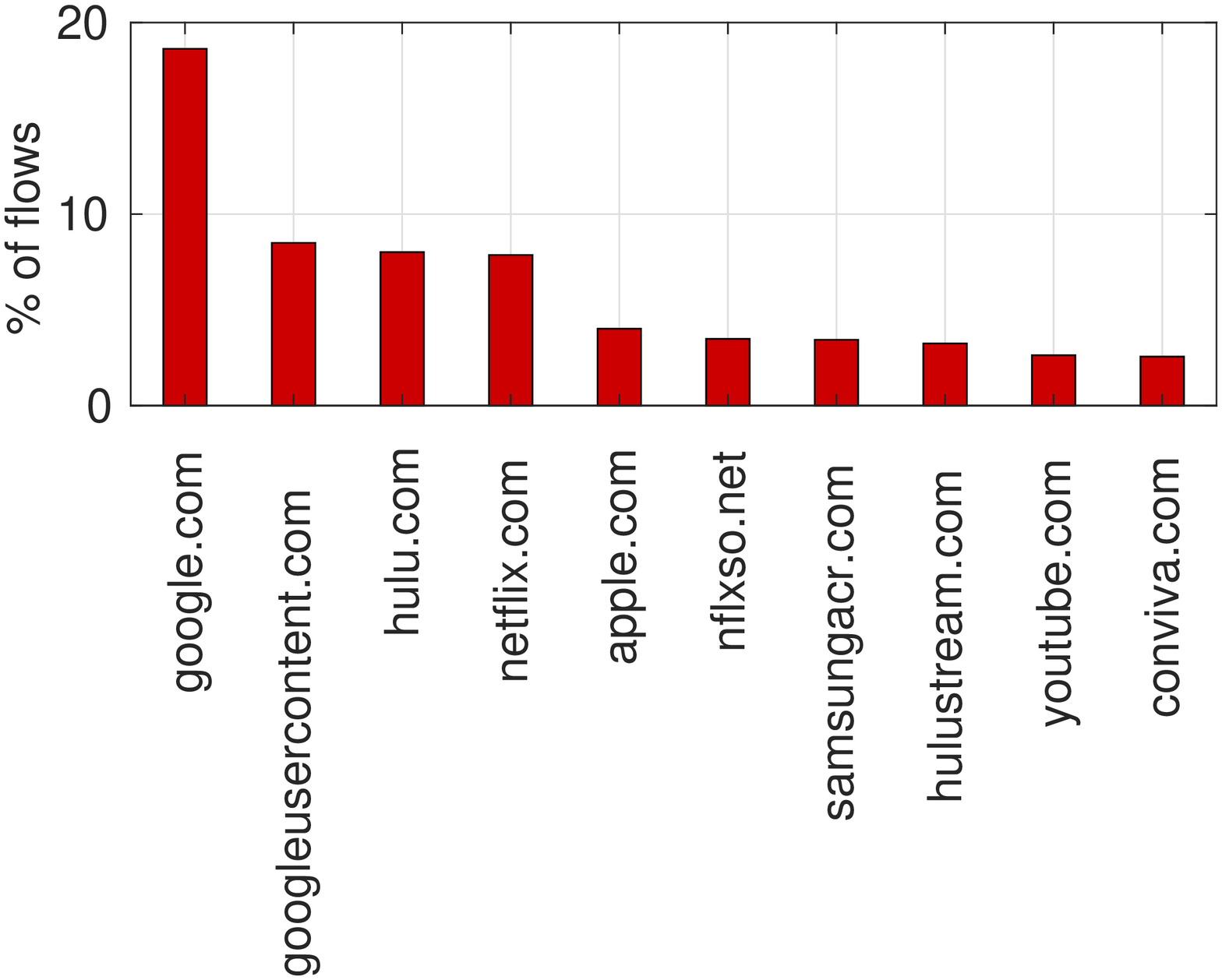}
    \caption{Smart TVs}
    \label{fig:smart_tv}
    \end{subfigure}
    \hfill
    \begin{subfigure}[t]{0.3\textwidth}
    \vskip 0pt
    \centering
    \includegraphics[width=\textwidth]{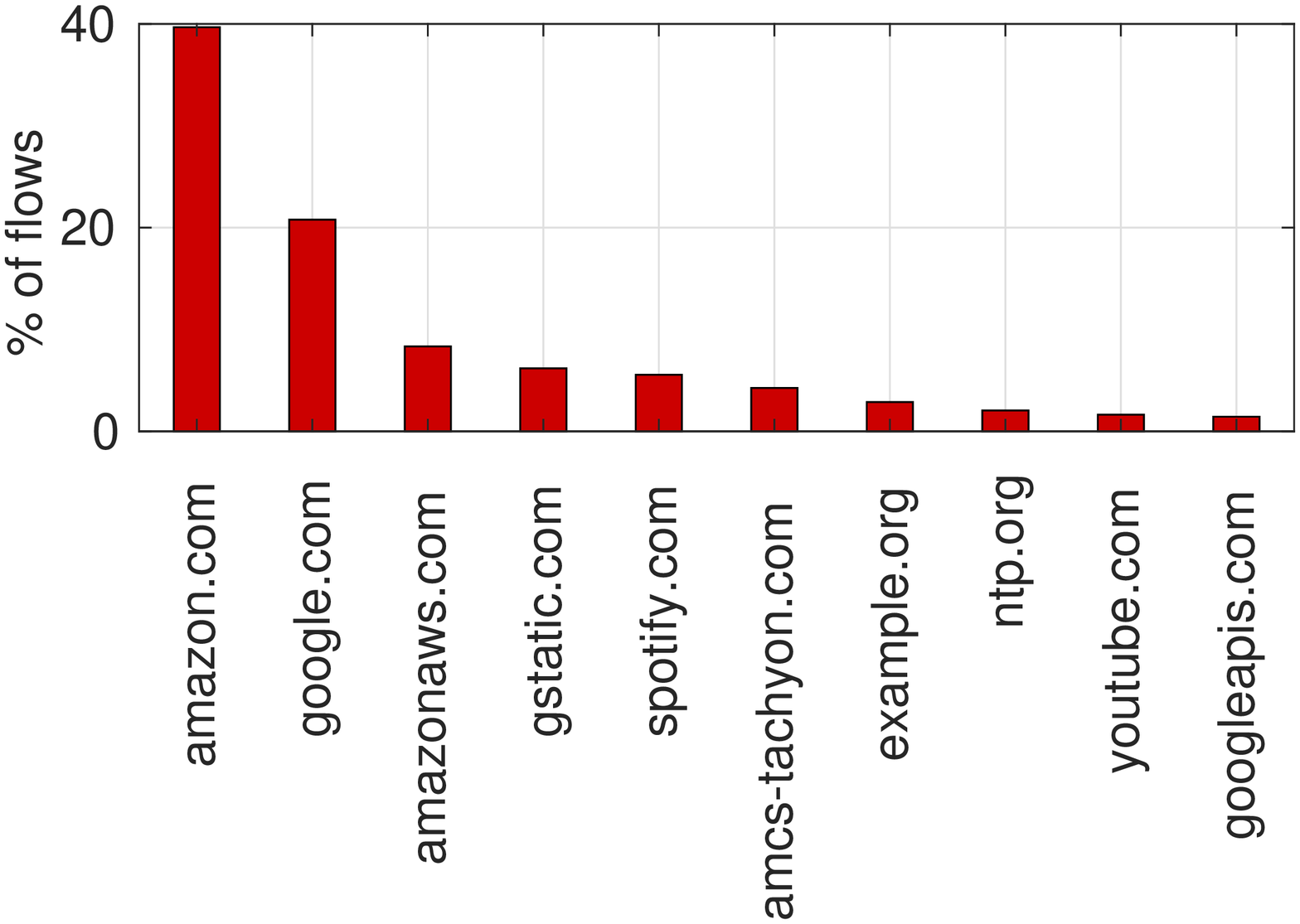}
    \captionsetup{skip=20pt}
    \caption{Smart Assistants}
    \label{fig:smart_assistant}
    \end{subfigure}
    \hfill
    \begin{subfigure}[t]{0.3\textwidth}
    \vskip 0pt
    \centering
    \includegraphics[width=\textwidth]{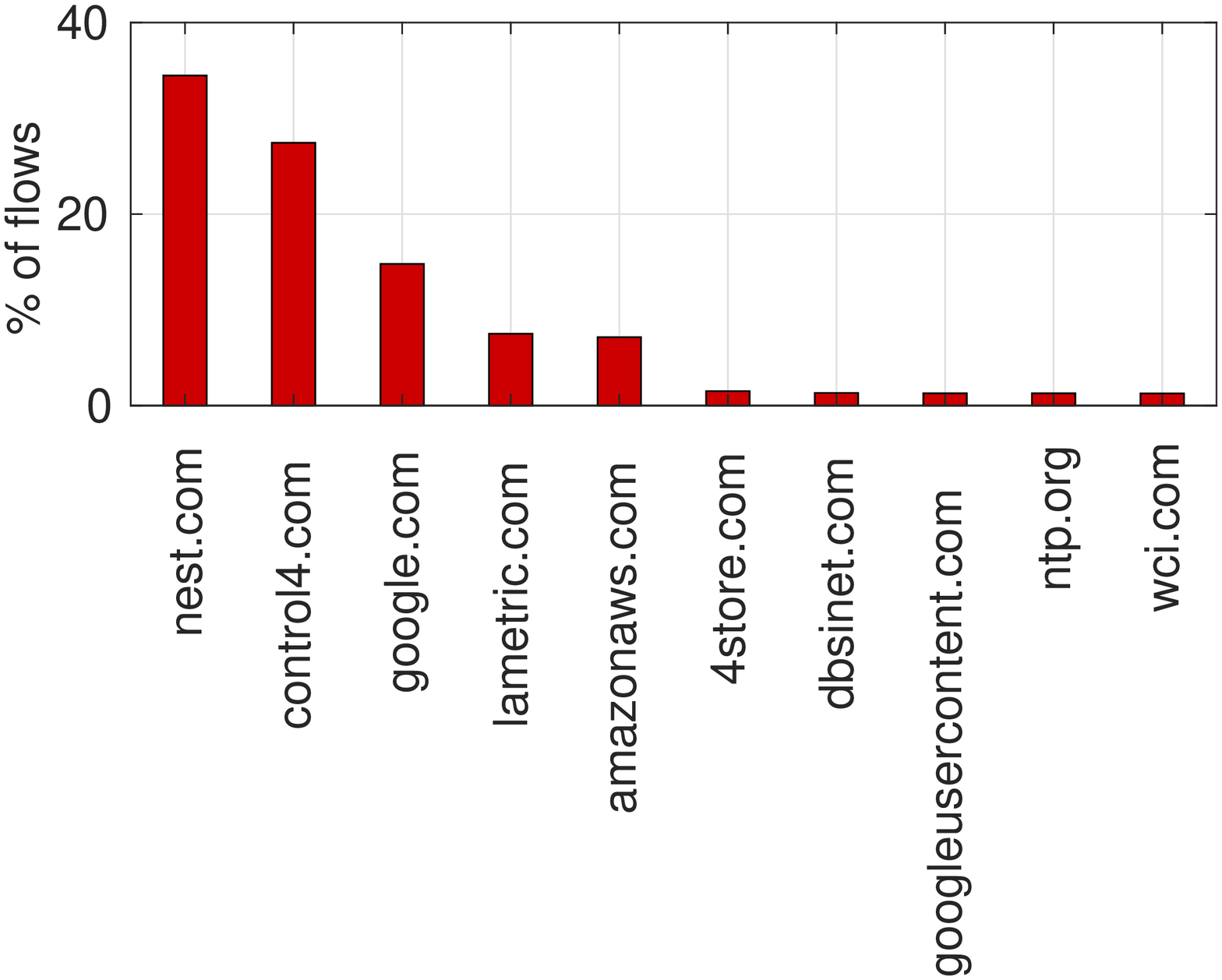}
    \caption{Home Automation}
    \label{fig:home_automation}
    \end{subfigure}
    \hfill
    \begin{subfigure}[t]{0.3\textwidth}
    \vskip 0pt
    \centering
    \includegraphics[width=\textwidth]{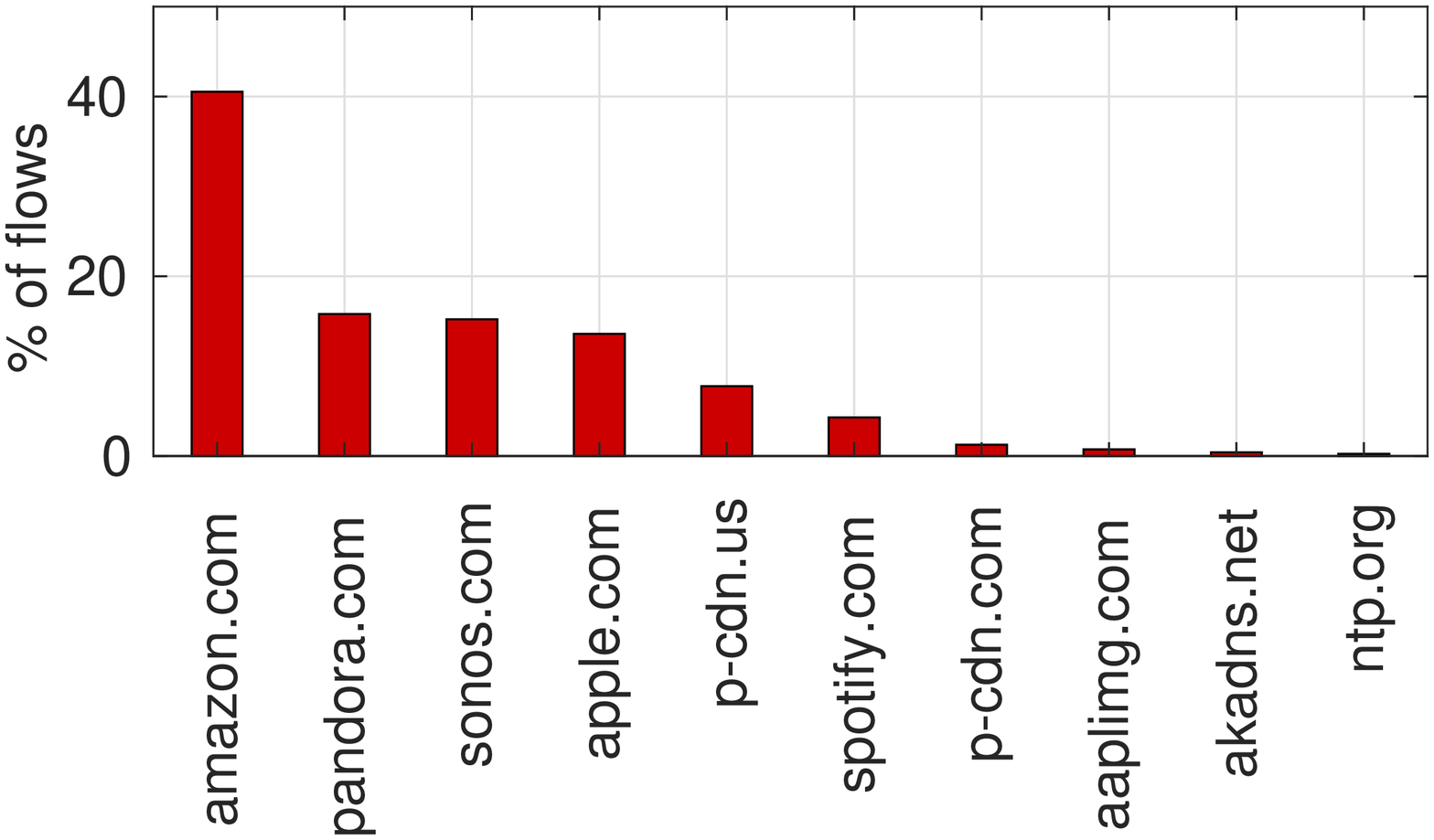}
    \captionsetup{skip=40pt}
    \caption{Smart Speakers}
    \label{fig:smart_speakers}
    \end{subfigure}
    \caption{Top 10 domains by flow ratio for selected smart home IoT device categories.}
    \label{fig:domain_plots}
\end{figure*}

\noindent\textbf{SLDs.}
We next analyze hostnames across different smart home IoT categories.
For simplicity, we map hostnames to Second Level Domain (SLD).
Figure \ref{fig:domain_plots} plots the top-10 SLDs for different smart home IoT categories.
We note that top-10 SLDs generally reflect device functionality.
For example, smart cameras connect to SLDs such as \textsf{xbcs.net} (owned by Belkin) to backup video footage, game consoles connect to gaming services such as \textsf{xboxlive.com}, and smart TVs connect to video streaming services such as \textsf{netflix.com}.
It is interesting to note that game consoles also accessed video streaming services indicating their dual-use as media streaming devices.
For smart TVs, along with video streaming SLDs we also observe \textsf{samsungacr.com}, which is associated with Samsung's Automatic Content Recognition (ACR) service.
ACR services are used to track users' viewing behavior on smart TVs and leveraged for ad targeting \cite{samsung2019acr,wolk2018forbesacr}.
Some smart home IoT devices periodically send `heartbeat' traffic to SLDs owned by their manufacturers such as  \textsf{lametric.com} (smart clock), \textsf{control4.com} (home automation), and \textsf{sonos.com} (smart speaker).

\noindent \textbf{\textit{Takeaway.}}
\textit{Smart home IoT devices communicate with services that are centralized on major cloud providers, which are adopted due to cost and efficiency for device manufacturers.
These services are tied with device functionality, such as gaming services for game consoles and media streaming services for smart TVs, control services for home automation devices and smart assistants.
Furthermore, there is interest from smart TV manufacturers to leverage their devices to track user behavior for advertising and tracking.
}

\section{Security \& Privacy Issues in Smart Home IoT}\label{sec:case_studies}
In this section, we investigate smart home IoT traffic with respect to specific cases.
These cases primarily highlight security and privacy concerns that arise with the proliferation of smart home IoT.

\subsection{Securing Smart Home IoT via Internet Access Control}
As smart home IoT devices and IoT in general become more ubiquitous, concerns have been raised with regards to how network access by such devices be controlled to prevent security issues such as device compromise.
Manufacturer Usage Description \cite{lear2019mud} (MUD) is a recently approved IETF standard (RFC 8520) that provides a standardized method for smart home IoT device manufacturers to specify the ports, protocols and network hosts that their devices will communicate with.
These MUDs can then be used by network administrators or gateway routers to develop Internet Access Control Lists (ACLs) to firewall smart home IoT devices to improve their security posture.
Researchers have built tools that can generate MUDs for devices given traffic traces \cite{hamza2018mud} to facilitate manufacturers, and utilized MUDs to propose methods for detecting attacks on smart home IoT devices \cite{hamza2019sosr}.
Furthermore, industry is also providing tools for manufacturers and network administrators to incorporate MUD-based IoT device management \cite{su2019ciscomud}.

\begin{figure}[!ht]
    \centering
    \includegraphics[width=\columnwidth]{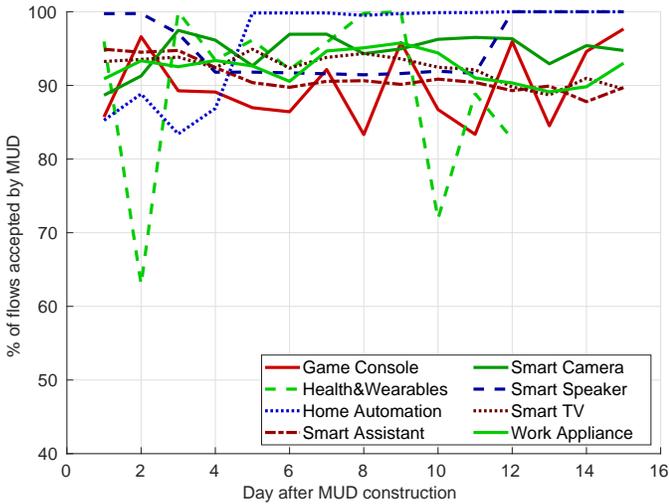}
    \caption{Average percentage of flows allowed by MUDs generated from device traffic from the first 72 hours per smart home IoT category.}
    \label{fig:mud_pass}
\end{figure}
These proposals and tools rely on the ability of device manufacturers to define MUDs that describe any legitimate traffic generated by smart home IoT devices.
MUDs for devices with well-defined functionality such as smart cameras and smart thermostats would be fairly easy to define.
However, MUDs for devices such as game consoles and smart TVs which access hosts not under manufacturer control may be difficult to define.
We illustrate this issue by evaluating the effectiveness of MUDs in-the-wild through analysis of traffic in our dataset.
Since MUDs are not currently deployed by manufacturers, we generate them by adapting MUD\textit{gee} \cite{hamza2018mud}.
For every smart home IoT device in our dataset, we generate a MUD using MUD\textit{gee}'s methodology over first 72 hours of traffic data for the the device.
We then test the MUD over that device's subsequent traffic, noting the amount of flows that would have been passed the MUD-based ACL for the device.
We plot the results of this test in Figure \ref{fig:mud_pass}, which shows the average percentage of flows that would have passed the ACL over the course of multiple days.
%

We observe high acceptance rates across all device categories, with smart speakers and home automation achieving 100\% acceptance for long periods of time.
Smart assistants, smart TVs and game consoles achieved high acceptance rates that fluctuated between between 85-95\%.
Heath \& wearables saw two days where acceptance rates fell to 64\% and 72\%.  
These drops happened due to Apple Watch devices, which accessed hostnames not observed during the traffic used for MUD generation.
These results illustrate that while MUDs can aid network administrators in developing solutions to secure smart home IoT devices through Internet access control, they require further work on how they are generated to account for cases where legitimate traffic is not accounted for in the MUD.

\subsection{Advertising and Tracking in Smart Home IoT}
Across the smart home IoT ecosystem, smart TVs have been found to track user behavior for targeted advertising, which for some manufacturers has become their main revenue stream \cite{niu2018roku,samsung2019acr}.
This has become a serious regulatory concern because smart TV users are tracked without their knowledge and consent. 
For example, Vizio was fined by the Federal Trade Commission (FTC) for collecting channel viewing history of users using ACR without user consent \cite{federal2017vizio}.
Recall from Section \ref{sec:who} that we observed the presence of \textsf{samsungacr.com} in smart TV traffic.
Recent research has also highlighted the prevalence of tracking in smart TVs \cite{moghaddam2019watching}.
We surmise that smart home IoT devices in general can be leveraged for tracking user behavior by manufacturers and third-parties whose services are accessed through these IoT devices.
Our goal in this section is to determine whether such tracking already exists in smart home IoT devices, including but not limited to smart TVs.

\begin{table}[t]
    \centering
    \begin{tabular}{p{2.3cm} R{1.6cm} R{1.6cm} R{1.3cm}}
         \textbf{Category}&\textbf{Total unique hosts}&\textbf{Found in Pi-hole list}&\textbf{\% of total} \\\hline
         Game Consoles & 32,259 & 992 & 3.1\% \\
         Smart TV & 9,684 & 576 & 5.9\% \\
         Smart Assistant & 2,091 & 48 & 2.9\% \\
         Smart Camera & 708 & 0 & 0\% \\
         Health \& Wearables & 257 & 5 &1.9\%\\
         Home Automation & 185 & 0 &0\%\\
         Smart Speaker & 184 & 1 &0.5\%\\
         Work Appliance & 37 & 1 &2.7\%\\\hline
         Smartphones & 65,625 & 2,796 &4.3\%\\\hline

    \end{tabular}
    \caption{The number and percentage of hosts detected by Pi-Hole as associated with ad/tracking.}
    \label{tab:track_ratio}
\end{table}

To this end, we use Pi-hole which is a tool for blocking advertisers and trackers across the whole network by monitoring DNS queries for hostnames and domains associated with them.
We use the default set of lists available in Pi-hole \cite{list2019pihole} to check the hostnames accessed by the devices in our dataset and count the number of hostnames that were found in the lists.
We count the total number of unique hostnames for each smart home IoT device category and the number of hostnames that were found on Pi-Hole's lists in Table \ref{tab:track_ratio}.
This allows us to understand the prevalence of advertising and tracking in smart home IoT.
We also count such hostnames for smartphones as a baseline comparison.

We note that 6 out of 8 smart home IoT categories accessed an ad or tracker hostname as marked by Pi-hole, indicating that some form of tracking or ad delivery is present in these categories.
Smart TVs communicated the most with ads and tracker hostnames at 5.9\% of all hostnames accessed by such devices.
Next, game consoles, smart assistants, and health \& wearable devices had 3.1\%, 2.9\% and 1.9\% of their hostnames marked as an ad or tracker.
Smart speakers and work appliances only accessed a singular ads/tracking hostname, which were \textsf{msmetrics.ws.sonos.com} and a \textsf{google-analytics.com} hostname respectively.
Note that while the Pi-hole list may capture many advertising and tracking services \cite{razaghpanah2018apps}, it may miss others that are unique to smart home IoT ecosystem \cite{moghaddam2019watching}.

For 3 smart home IoT categories (game consoles, smart TVs and smart assistants) that accessed more than 10 ads/tracking hostnames, we extract the domains of such hostnames and determine the top 15 domains with respect to the number of devices they were accessed by.
We first note that the list of top 15 domains is very similar across these devices, with most ad/tracking hosts originating from Google owned domains, such as \textsf{doubleclick.com} and \textsf{googlesyndication.com}.
This indicates the capability of Google to possibly track user behavior in some form on smart home IoT.
Other tracking domains include \textsf{imrworldwide.com} (owned by Nielsen Online), \textsf{casalemedia.com} (owned by Casale Media) and \textsf{invitemedia.com} (owned by Invite Media), which would also gain the ability to track user behavior on smart home IoT devices.

The presence of these hostnames is indicative of the fact that tracking has reached smart home IoT devices.
To mitigate such tracking, users may use network-level blocking solutions such as Pi-hole \cite{pihole2019} which block DNS requests for advertising and tracking services using block lists.
However these block lists, which are manually curated based on informal crowdsourced user reports, are prone to mistakes and trivial circumvention by advertisers and trackers \cite{Alrizah2019errors}.

\begin{figure}
    \centering
    \includegraphics[width=\columnwidth]{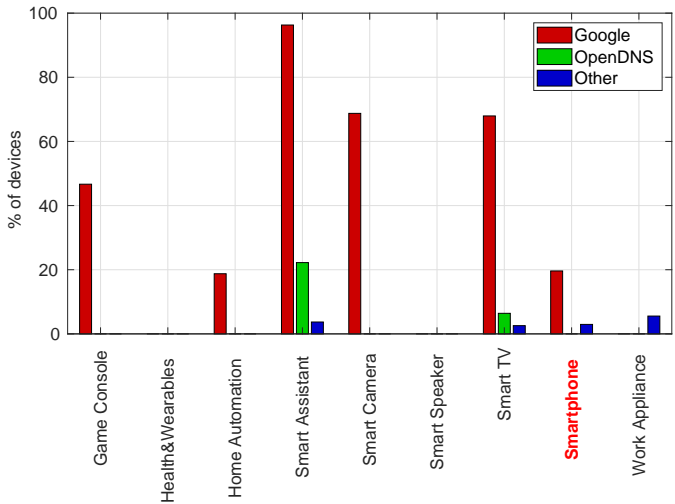}
    \caption{Percentage of devices that accessed public DNS servers across smart home IoT categories. Google DNS servers and OpenDNS servers were most prevalent.}
    \label{fig:dns_plot}
\end{figure}

\subsection{Use of Public DNS by Smart Home IoT}
Our gateways are instrumented to run a DNS server that is assigned via DHCP and is responsible for answering DNS queries sent by local network devices.
However, devices may be configured to use hard-coded public DNS servers.
We analyze device traffic data to determine the prevalence of this practice across smart home IoT devices.
Our gateways do not log flows to the local DNS server hosted on it, but logs DNS queries to public DNS servers as UDP or TCP flows on port 53.
We plot the percentage of devices which accessed an public DNS server for each smart home IoT device category with smartphones as baseline in Figure \ref{fig:dns_plot}.
All smart home IoT categories except health \& wearables and smart speakers access had devices which accessed an public DNS server.
Smart assistants were the most prevalent, with 98\% of devices accessing Google DNS servers including all Google Home devices.
Google DNS was also popular amongst smart cameras, smart TVs and game consoles with 68\%, 68\% and 46\% of such devices accessing it.
We also note that OpenDNS servers were accessed by smart assistants and smart TVs by Amazon-manufactured devices.

Devices may choose to use hard-coded public DNS servers due to various reasons.
For instance, the Google Chromecast is hard-coded with Google DNS server addresses to prevent access to geo-locked content on services such as Netflix and Hulu \cite{bartholomew2016chromecastdns}.
Such behavior has also been noted in recent work \cite{huang2019iot} where Netflix hostnames were exclusively resolved through Google DNS on Roku-based smart TVs.
Other reasons include preemptively avoiding problems caused by mismanaged DNS servers hosted by ISPs \cite{bort2019chromecastdns}, which leads to users blaming the device for the problems.
While such reasons may be valid, they take away control from the user on how devices on their networks communicate with the Internet.
Furthermore, the use of hard-coded DNS also renders network-level blocking solutions \cite{pihole2019,adguard2019} invalid as they would not be used to resolve DNS queries.

\begin{figure}
    \centering
    \includegraphics[width=\columnwidth]{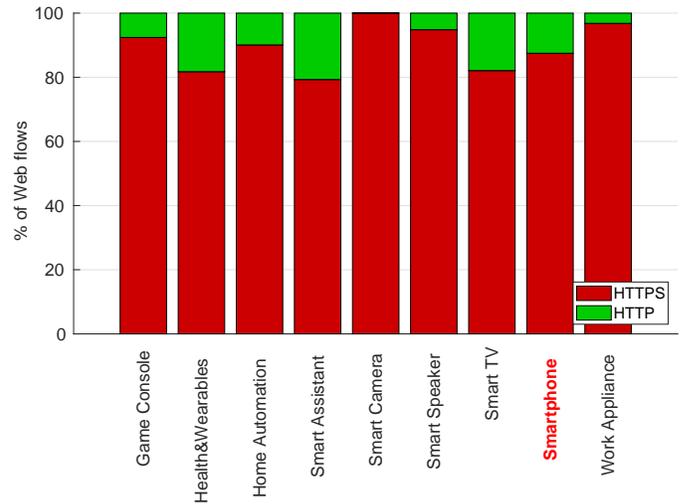}
    \caption{Percentage of Web traffic flows sent over HTTP or HTTPS by smart home IoT devices.}
    \label{fig:http_plot}
\end{figure}
\subsection{Prevalence of Unencrypted Traffic in Smart Home IoT}
Many smart home IoT devices are designed to access the Web for various services, which may be available via HTTP or HTTPS.
Since HTTP traffic is typically not encrypted, access to services over HTTP can leak sensitive information about the user to a passive network observer.
To this end, we explore the proportion of Web traffic for smart home IoT categories that is accessed through HTTP and HTTPS by analyzing traffic destined for port 80 and port 443 for either protocol respectively in Figure \ref{fig:http_plot} with smartphones as baseline.
We note that all smart home IoT devices except smart cameras access the Web over HTTP for some portion of traffic, with health \& wearables, smart assistants, and smart TV generating around 20\% of their traffic over HTTP, more than that accessed by smartphones.
A closer inspection of flow data for these categories reveals that such flows are mostly associated with ads and tracking SLDs such as \textsf{scorecardresearch.com} and \textsf{imrworldwide.com} across all three categories.
Health \& wearables also accessed SLDs such as \textsf{fitbit.com} which relate to device functionality, while smart assistants and smart TVs accessed media streaming SLDs for services such as Netflix, Hulu and Spotify.

There may be various reasons for services to be still provided over HTTP.
As noted by Englehardt and Narayanan \cite{englehardt2016online}, services may be hesitant to move to HTTPS if they use any third-party resources that are HTTP-only.
These resources are typically ads and trackers, which were also found to be predominantly over HTTP in our own analysis.
Hill and Mattu \cite{hill2018gizmodo} noted in their study that smart TVs sent information on use of Hulu services to tracking hostnames, leaking information about user viewing behaviors.
Smart home IoT devices may also access services over HTTP due to limitations in the device itself.
For instance, \textsf{fitbit.com} was accessed over HTTP by a Fitbit Aria smart scale that is no longer supported by the manufacturer with firmware updates, with the last update being a security patch in 2016 \cite{nichols2016fitbit}.
While the current Fitbit API \cite{fitbit2019api} is restricted to HTTPS only, it is likely that HTTP support is retained for backwards compatibility.
A key concern with smart home IoT is how devices that are no longer supported by the manufacturer are handled with regard to issues such as reliability and security.
As proposed by Fagan et al. \cite{fagan2019core}, features for IoT devices should be designed to account for their lifespan and as such manufacturers should ensure that their devices can be updated to maintain sufficient reliability and security.

\section{Related Work}\label{sec:related_work}
The growth of smart home IoT devices has brought interest to such devices by malicious actors as a viable target.
One famous instance is the Mirai botnet\cite{antonakakis17usenixsec} composed mostly of smart cameras, used in Distributed Denial of Service (DDoS) attacks against Dyn and KrebsOnSecurity\cite{krebsMirai}.
Recognizing this threat, researchers have focused on not only studying smart home IoT devices from the lens of security and privacy, but also understanding how they are being adopted by consumers.
We discuss some of this prior work here.

\vspace{.05in} \noindent\textbf{Smart home IoT in testbed environments.}
Much of initial work on understanding smart home IoT device behavior through passive network observation at the home gateway \cite{amar2018arxiv,apthrope2016dat,hill2018gizmodo,shahid2018bigdata, wood2017iotsnp} focuses on understanding the implications of such behavior from a user privacy perspective.
More recently, Ren et al. \cite{ren2019imc} conducted experiments on smart home IoT device behavior on 81 devices spread across an US-based and an UK-based testbed environment.
Their experiments showed that smart home IoT devices in their dataset routinely expose information to eavesdroppers via plaintext flows or to destinations not owned by manufacturers, and routinely communicate with destinations outside their privacy jurisdictions.
Note that we also found cases of possible information exposure via HTTP flows as well as connections to ad and tracking hostnames.
These studies leverage the home gateway as a vantage point, which is able to provide fine-grained insights into device behavior.
However these studies are limited by their use of \textit{testbed environments and their selections of IoT devices}, which cannot be considered representative insights for all smart home IoT devices.

\vspace{.05in} \noindent\textbf{Tools for studying smart home IoT at scale.}
Researchers have aimed to build tools that allow them to collect data from smart home IoT devices on large scales.
These tools are designed to collect data through in-path passive monitoring of network traffic, or through off-path active probing of devices to collect responses.
Huang et al. designed IoT-Inspector \cite{huang2019iot} as an in-path tool designed to collect crowd-sourced information on smart home IoT device behavior in the wild, using Address Resolution Protocol (ARP) spoofing to capture network traffic generated by smart home IoT devices. 
IoT-Inspector is primarily targeted towards users looking to understanding how their smart home IoT devices communicate with the Internet.
Based on data collected from 8,131 devices, Huang et al. find devices that communicated over HTTP and used weak cipher suites for Transport Layer Security (TLS).
Furthermore they also found smart TVs in their dataset to connect to advertising and tracking domans, as well as use hard-coded DNS servers, both of which we also show in our case study analysis.
While IoT-Inspector collects data passively it does so when a user \textit{initiates it, which limits the ability of this data to reflect trends in smart home IoT behavior.}
Work has also focused on studying smart home IoT through Internet-scale measurements.
To this end, tools such as Internet-wide active scanners of network hosts have been leveraged for such work.
Shodan \cite{matherly2017shodan} is a search engine developed to identify IoT devices using probe traffic to known ports for services such as HTTP/HTTPS, SSH and FTP.
Similarly, Censys \cite{durumeric15ccs} also provides internet-wide scanning for services and devices but also supports crowd-sourced annotation of device information.
Such services have been used to search the Internet for smart home IoT devices which are compromised by malware \cite{antonakakis17usenixsec,herwig18ndss}.
Active probing measurements only provide information on how IoT devices respond to them, providing \textit{no insight on passively observed behavior}.
\vspace{.05in} \noindent\textbf{In the wild measurements.}
There has been prior work on how smart home IoT devices or Internet-connected devices behave in-the-wild i.e. when they are used by normal users in their homes. 
Hill and Mattu \cite{hill2018gizmodo} conducted a 2-month study on smart home IoT devices placed in Hill's home.
They studied how traffic behavior from certain devices can be used to infer user behavior and preferences, and how this information may be leveraged by third-parties.
Unfortunately, their insights may not be representative of general smart home IoT behavior in the wild given the sample size of only one home.
Grover et al. \cite{grover2013imc, sundaresan2014usenix} studied home networks in 100 homes across 21 countries via deployed routers instrumented with custom firmware to conduct active and passive measurements.
They highlight differences between homes in developing and developed countries through the lens of the availability, infrastructure, and usage patterns of home networks.
They note that home networks in developing countries experience more Internet interruptions, but are similar to home networks in developed countries in terms of the number of connected devices.
They also analyze traffic data from 25 houses to observe usage patterns. 
This work is mainly limited to studying network performance in home networks and does provide insight into the behaviors of individual devices including IoT. 
%
%
More recently, Kumar et al. \cite{kumar18usesec} presented an active measurement study of 83 million devices in 16 million home networks around the world.
Their analysis primarily focused on the presence of various IoT device types on home networks, noting that significant amounts of homes in North America, Western Europe and Oceania have at least one IoT device present.
They also note that many IoT devices still exhibit bad security posture through the exposure of services, such as FTP and Telnet, or the use of default credentials in administration interfaces.
While this work provides a valuable large-scale survey of different IoT devices, it does not passively capture behavior and usage characteristics of smart home IoT devices in the wild.

Our work advances the research by conducting passive measurement and in-depth behavioral characterization of a diverse set of smart home IoT devices in the wild.
As we discuss next, we highlight several new and interesting characteristics of the smart home IoT ecosystem that warrant further research.




%
%

\section{Conclusion and Discussion}\label{sec:conclusion}
In this paper, we presented a characterization of smart home IoT traffic in the wild.
We deployed instrumented home gateways to gather and analyze network traffic logs from more than 200 homes containing a wide variety of IoT devices. 
As we discuss next, our characterization of different aspects of smart home IoT traffic uncovers several interesting findings that warrant future investigation.

We find that device functionality clearly influences smart home IoT traffic---devices that access media over the Web exhibit high-volume diurnal traffic that matches human activity patterns while devices that provide automation functionalities exhibit low-volume traffic with sub-hour periodicity.
These findings show that IoT traffic patterns can be leveraged to not only improve device identification approaches   \cite{ortiz2019iotdi} but also assess the effectiveness of IoT device activity fingerprinting \cite{apthrope2016dat}, where user activities may be inferred through analysis of smart home IoT traffic.
Our insights can also help in developing better methods to evade IoT device activity fingerprinting through traffic shaping techniques \cite{apthorpe2019pets}.

We also find that smart home IoT traffic reflects significant centralization towards major cloud providers and public DNS providers.
While centralization of the cloud brings benefits such as higher availability, redundancy, and ease of implementation, it also brings risks due to monopolization as well as the possibility of malicious intentions (e.g. censorship, surveillance) by the cloud provider.
Multi-cloud solutions to address these concerns caused by relying on a single cloud provider are an active research area \cite{alzain2012cloud,tato2018koala,yan2019gecko}. 
Centralization of DNS also brings its own dangers by presenting a single point of failure, as evident from the Dyn DDoS attack \cite{bates2018evidence}.
Devices with hard-coded DNS servers could cease to function if the DNS server is down or could be compromised if the DNS server is compromised.
Device manufacturers should ensure that their devices are designed with suitable countermeasures to prevent such failures.

Our findings also raise privacy concerns by providing evidence of unencrypted traffic over HTTP.
To prevent leakage of personal information through unencrypted traffic, prior work has investigated using Virtual Private Network (VPN) at the home gateway to encrypted and wrap traffic into a single flow between source and destination IP addresses of VPN endpoints \cite{apthorpe2019pets,nordvpn2019,shif2018improvement}. 
Unfortunately, VPNs only prevent eavesdropping of unencrypted network traffic from an adversary at the access ISP but not beyond the external VPN endpoint \cite{apthorpe2019pets}.
Furthermore, using a VPN comes with a performance penalty as the traffic is first routed to VPN servers before being sent to the actual destination.
Recent work \cite{varvello2019vpn0,prince2019warp} has focused on improving VPN performance while maintaining the security and privacy guarantees provided by them.

We also find prevalence of third-party advertising and tracking services in smart home IoT traffic. 
To prevent tracking from smart home IoT devices, users can deploy network-level blocking tools such as Pi-hole \cite{pihole2019}.
However, existing network-level blocking tools are mainly geared towards web and mobile, and are known to suffer from significant blind spots for smart home IoT traffic \cite{moghaddam2019watching}.
Moreover, it is inherently challenging for network-level blocking to block first-party tracking \cite{cimpanu2019dropadblock}.
Our work highlights the need for further research to improve the effectiveness of network-level blocking tools for smart home IoT traffic.



%
\section{Acknowlegments}
This work is supported in part by the National Science Foundation under grant numbers 1815131 and 1617288, and by Minim.
The authors would also like to thank the team at Minim for their help with collecting and analyzing smart home IoT traffic.

\bibliographystyle{abbrv}
\bibliography{references}

\end{document}